\documentclass{PoS} % for review and submission

\usepackage{graphicx}  % needed for figures
\usepackage{dcolumn}   % needed for some tables
\usepackage{bm}        % for math
\usepackage{amssymb}   % for math
\usepackage{amsmath}
\usepackage{cancel}
\usepackage{enumerate}

\usepackage{epstopdf}
\newcommand{\bee}{\begin{equation}}
\newcommand{\ee}{\end{equation}}
\newcommand{\beea}{\begin{eqnarray}}
\newcommand{\eea}{\end{eqnarray}}

\newcommand{\la}{\ensuremath{\lambda} }

\newcommand{\X}{\ensuremath{\!\times\!} }

\newcommand{\Sb}{\ensuremath{\cancel{S^4}} }

\newcommand{\refcite}[1]{Ref.~\cite{#1}}
\newcommand{\eqn}[1]{Eq.~\ref{#1}}
\newcommand{\fig}[1]{Fig.~\ref{#1}}
\title{Mass anomalous dimension from Dirac eigenmode scaling in conformal and confining systems}

\ShortTitle{Mass anomalous dimension from Dirac eigenmode scaling}

\author{\speaker{Anna Hasenfratz}, Anqi Cheng, Gregory Petropoulos, David Schaich\\
        Department of Physics, University of Colorado, Boulder, CO-80309-390\\
        E-mail: \email{anna@eotvos.colorado.edu}}

\abstract{The mode number of the Dirac operator scales with an exponent related to the mass anomalous dimension $\gamma_m$.  This relation holds both in IR-conformal systems, as well as in confining systems for large enough eigenvalues.
We investigate the $N_f=4$, 8 and 12 flavor SU(3) systems at several couplings near the chiral limit, and show that in general the scaling exponent varies with the eigenvalue, describing the dependence of $\gamma_m$ on the energy (or, equivalently, on the running coupling).  This energy dependence can be explored even with fixed lattice parameters (bare coupling and mass).  We find that for the 4 flavor system the mass anomalous dimension decreases as the energy increases, consistent with perturbative expectations.  For the 8 flavor system the energy dependence is too weak to be observable at present. The 12 flavor system at our strongest couplings shows the anomalous dimension increasing with energy, consistent with backward flow and the presence of an infrared fixed point.  At weaker couplings we determine a preliminary value for the mass anomalous dimension of the 12 flavor system at the infrared fixed point, $\gamma_m^* = 0.27(3)$.}

\FullConference{30th International Symposium on Lattice Field Theory \\ 24--29 June 2012 \\ Cairns, Australia}

\begin{document}
\section{Introduction} % Draft complete

The lattice study of gauge systems with many flavors or higher representation fermions is motivated in part by the possibility of electroweak symmetry breaking via new strong dynamics. The discovery of a 125GeV Higgs-like particle puts severe restrictions on many BSM scenarios, including models of new strong dynamics, which are now required to predict a light composite Higgs. Even if the particle discovered at 125GeV turns out to be an elementary scalar, the study of these strongly coupled systems remains important not only for theory but model building as well.

While the numerical simulation of many-fermion systems is not fundamentally different from well understood QCD simulations, it is increasingly clear that QCD-like analyses are not always optimal.  The large number of fermions requires working with relatively strong couplings where unusual phases, unexpected lattice artifacts and spurious UV fixed points can influence the critical behavior \cite{Cheng:2011ic}. We use several complementary techniques and approaches to investigate these lattice models, comparing and contrasting systems with different numbers of fermions to distinguish conformal and chirally broken, confining behavior. In this paper we consider the scaling of the Dirac eigenmodes; our work on the phase structure, finite temperature behavior, and Monte Carlo renormalization group (MCRG) analysis is discussed in other contributions to these proceedings~\cite{Schaich:2012lat12, Petropoulos:2012Lat12}.

Our gauge action consists of fundamental and adjoint plaquette terms and we use nHYP smeared staggered fermions.
The details of the lattice action can be found in \refcite{Hasenfratz:2011xn}.  In this work we are interested in the weak coupling behavior of these systems, and we avoid the single-site shift symmetry broken ($\Sb$) lattice phase discussed in Refs.~\cite{Cheng:2011ic, Schaich:2012lat12}.
The $N_f=4$, 8, 12 and 16 flavor systems have been investigated extensively by several other groups; recent references include~\cite{Fodor:2012uu, Aoki:2012eq, Deuzeman:2012pv, Lin:2012iw} and earlier works are reviewed in \refcite{Neil:2012cb}.

\section{The eigenmode density and mode number}

The eigenmodes of the Dirac operator contain a wealth of information about the dynamics of lattice systems. When the infrared behavior is captured in a Random Matrix Theory (RMT) universality class, the distribution of the low eigenmodes predicts physical quantities like the chiral condensate $\Sigma$.
Conformal theories are more difficult to analyze as there are no RMT predictions for the individual eigenmodes.  One approach is analyze the low-lying eigenmodes on several different volumes; a simple finite-volume scaling fit predicts the mass anomalous dimension, though with large systematic errors~\cite{DeGrand:2009hu, Cheng:2011ic}.

It is more reliable to consider the eigenmode density $\rho(\lambda)$ or its integral, the mode number
\begin{equation}
  \label{eq:modenumber}
  \nu(\lambda) = V \int_{-\lambda}^{\lambda} \rho(\omega)d\omega
\end{equation}
where $V$ is the volume of the system.
The mode number is renormalization group invariant even at finite fermion mass,
% $ \nu_R(\lambda_R, m_R) = \nu(\lambda,m)$
and can be used to predict the chiral condensate in QCD-like systems with high precision \cite{Giusti:2008vb}.

In conformal systems where the low energy dynamics is governed by an infrared fixed point (IRFP), the general scaling form
$\rho(\lambda) \propto \lambda^{\alpha}$
predicts % the mode number for small eigenvalues as % How small does the eigenvalue have to be for this prediction to be reasonable?
\begin{equation}
  \label{eq:nu_scaling}
  \nu(\lambda) = c V \lambda^{\alpha+1} = c ( L \lambda^{(\alpha+1)/4})^4,
\end{equation}
where $L=V^{1/4}$ is the linear system size. Using the renormalization group invariance of the mode number, this relates the exponent $\alpha$ and the scaling dimension of the mass $y_m$
\begin{equation}
  \label{eq:ym_vs_alpha}
 y_m = 1 + \gamma_m = \frac{4}{\alpha+1}.
\end{equation}
\refcite{Patella:2012da} used the mode number, evaluated with a stochastic method \cite{Giusti:2008vb} in a wide eigenvalue range, to predict the mass anomalous dimension for SU(2) gauge theory with two adjoint fermions.
In this work we follow a similar approach both in chirally broken and IR-conformal theories.
%We consider $N_f=4$, 8 and 12 flavor staggered fermions coupled to SU(3) gauge field and investigate the dependence on the sea fermion mass and volume.

\section{The energy dependence of the scaling dimension}

The eigenvalues $\lambda$ of the Dirac operator have dimension of mass.  From $\la \sim 0$ to the cutoff scale $\la \sim a^{-1}$, the eigenvalue distribution reflects the properties of the system from the infrared to the ultraviolet.  In a chirally broken system we can distinguish three regions:
\begin{enumerate}[I.]
  \item Low energy region, below the chiral symmetry breaking scale. Here the eigenvalues describe the infrared behavior of the system, as is reflected in the Banks--Casher formula $\rho(0) = 2 \Sigma/\pi$ \cite{Banks:1979yr}. The correction in $\lambda$ can be calculated in chiral perturbation theory and the scaling of the mode number predicts the chiral condensate $\Sigma$.
  \item Above the chiral symmetry breaking scale but still well below the cutoff
the system is governed by the Gaussian fixed point and
the eigenvalue density scales according to \eqn{eq:nu_scaling} with an energy dependent scaling exponent.
 At weak coupling, perturbation theory predicts the one-loop universal running of the anomalous dimension as $\gamma_m = 6 C_F g^2/(4\pi)^2 + O(g^4)$ for fermions in the fundamental representation.
  \item The high energy, ultraviolet part of the spectrum for lattice systems is non-universal, dominated by discretization effects.
\end{enumerate}

%By tuning the bare coupling towards the Gausssian fixed point $g^2=0$ larger and larger part of region II can be investigated.

In chirally broken systems the anomalous dimension $\gamma_m$ depends on the running coupling and consequently the energy scale. This is reflected in the scaling form of \eqn{eq:nu_scaling} as a scaling exponent that depends on $\lambda$ as a measure of the energy scale.  Just above the chiral symmetry breaking scale $\gamma_m$ is large; in the standard folklore $\gamma_m \approx 1$, therefore $\alpha \approx 1$. At high energies perturbation theory predicts $\gamma_m \searrow 0$ and $\alpha \nearrow 3$. % Can just use \to if these arrows are too weird

Conformal systems have only regions II and III. The low energy end of the spectrum reflects the properties of the infrared fixed point and the scaling form \eqn{eq:nu_scaling} predicts the anomalous dimension at the IRFP, $\gamma_m = \gamma_m^*$.
Starting from a weak bare coupling, $\gamma_m$ at high energies will again approach the perturbative $\gamma_m = 0$. There is no perturbative prediction for what happens if the bare coupling is in the strong coupling side of the basin of attraction of the IRFP.\footnote{The location of the IRFP depends on the specific renormalization group transformation considered; it is not a physical observable, and the strong coupling side of the IRFP is not well-defined.  However it is possible to observe whether or not a lattice system is in the strong coupling side of the IRFP's basin of attraction.}

The above discussion might sound speculative at this point. In the following we will support it using data from our 4, 8 and 12 flavor simulations.
\newpage % Without this the footnote is shoved onto page 4 -- no idea why, since there's plenty of space for it on page 3

\section{Finite mass and finite volume effects}
%%%%%%%%%%%%%%%%%%%%%%%%%%%%%%%%%%
\begin{figure}[tb]
  \centering
  \includegraphics[width=0.45\linewidth]{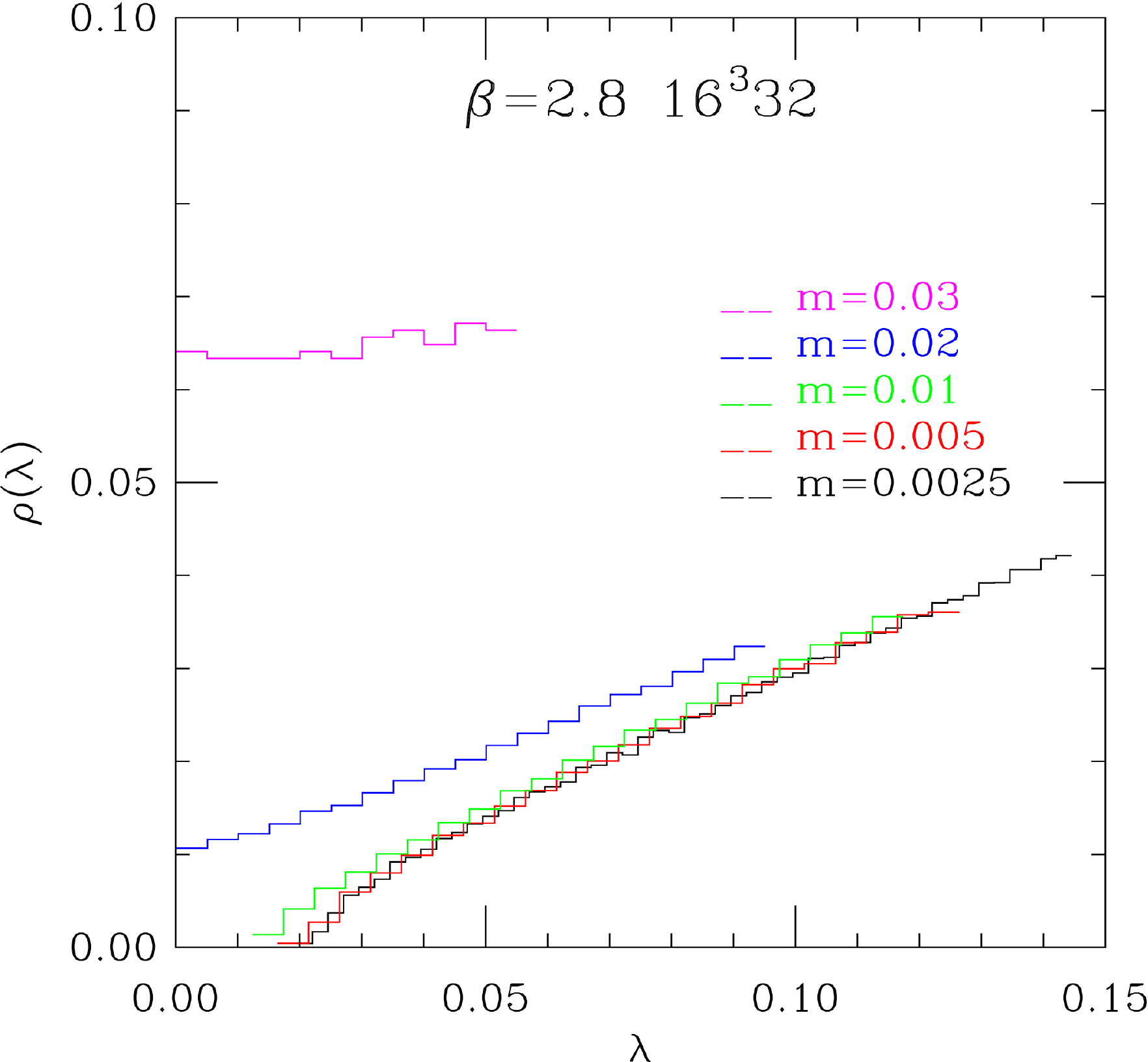}\hfill
  \includegraphics[width=0.45\linewidth]{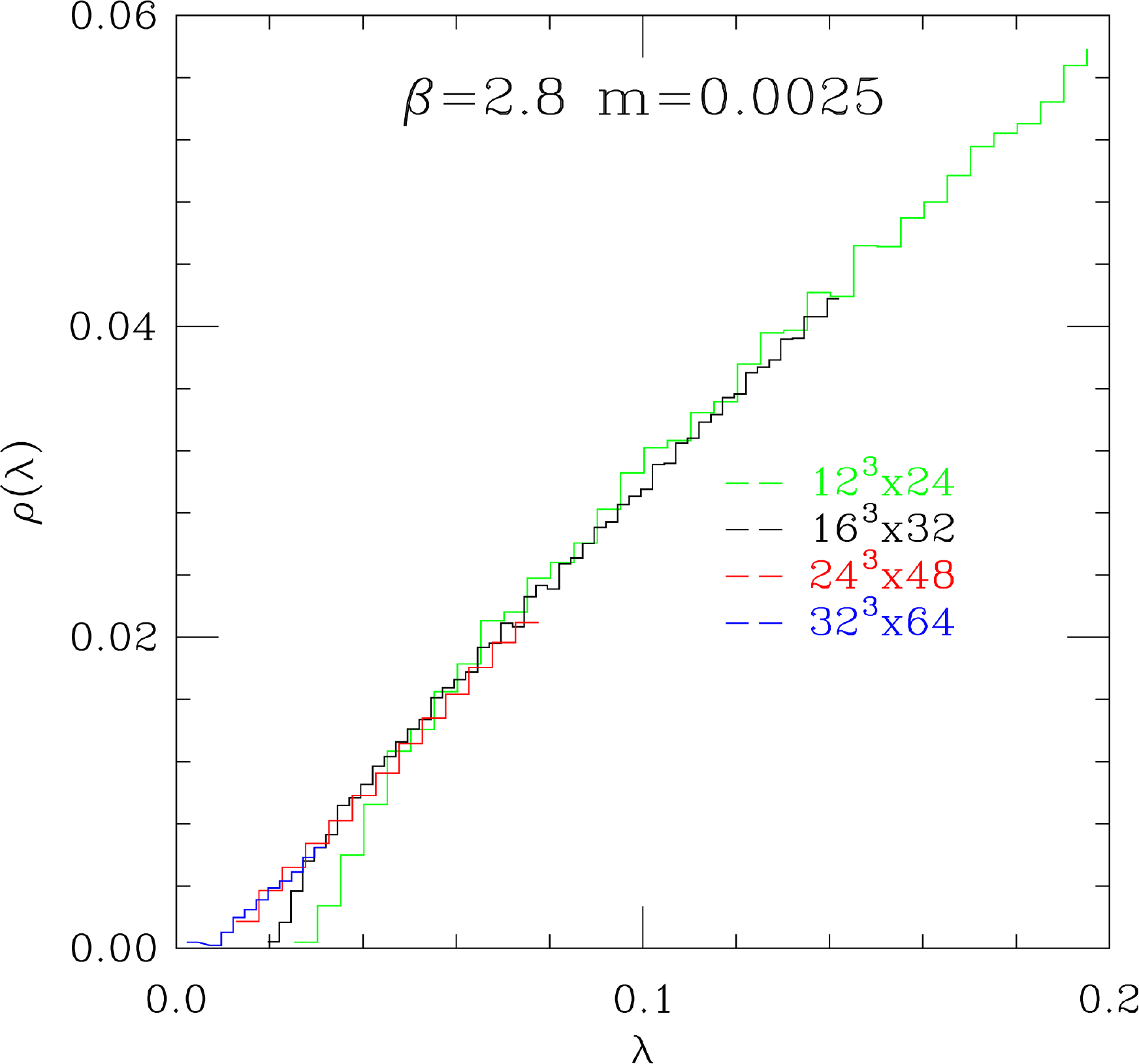}
  \caption{\label{fig:compare} The eigenvalue density $\rho(\lambda)$ at $\beta_F=2.8$ of the 12 flavor system. Left panel: $16^3\X32$ volumes at various sea fermion mass values. Right panel: mass $am=0.0025$ on various volumes.}
\end{figure}
%%%%%%%%%%%%%%%%%%%%%%%%%%%%%%%%%

The scaling form of \eqn{eq:nu_scaling} assumes infinite volume and vanishing fermion mass. In practice neither of these conditions are satisfied in lattice simulations.
The authors of Refs.~\cite{Giusti:2008vb,Patella:2012da} argue that the fermion mass affects only the low eigenmodes.
% even when the mode number of the massive hermitian Wilson fermion Dirac operator is considered.
Since we use staggered fermions we evaluate the eigenvalues of the massless Dirac operator, on configurations generated with nonzero sea fermion masses.
The left panel of \fig{fig:compare} shows $\rho(\lambda)$ for different sea fermion masses in the 12 flavor system at $\beta_F=2.8$ on $16^3\X32$ volumes. This coupling is safely on the weak coupling side of the \Sb phase \cite{Cheng:2011ic}. For $am \ge 0.02$, the eigenvalue density depends strongly on the mass, and appears unlikely to become mass independent even at larger $\lambda$.  On the other hand for $am\le0.01$ the mass dependence rapidly disappears as $\lambda$ increases, suggesting that here it is possible to reach the chiral limit by simple extrapolation.

The right panel of \fig{fig:compare} illustrates the volume dependence of $\rho(\lambda)$ for the smallest mass in the left panel, $am=0.0025$.
The system is chirally symmetric on all four volumes considered, with $\rho(0) = 0$.  At the small $\lambda$ where the density becomes nonzero, there is transient volume dependence.  All four volumes produce consistent results for larger $\lambda$ that are still well below the cutoff scale.
While there remains a small volume dependence even in this regime, an infinite volume extrapolation is feasible.

Even in systems known to exhibit spontaneous chiral symmetry breaking we find consistent, largely volume independent $\rho(\lambda)$ for $\lambda$ above the chiral symmetry breaking scale.  By comparing data on different volumes and at different sea fermion masses it is possible to identify the regime where both the infinite volume and chiral extrapolations are feasible.

\section{Scaling of the mode number}
In this section we present results for the mass anomalous dimension based on the scaling of the mode number (\eqn{eq:nu_scaling}) within region II.
We use $12^3 \X 24$, $16^3 \X 32$ and $24^3 \X48$ volumes, generate ensembles of 50--100 thermalized configurations, and calculate 300--1000 eigenvalues on each configuration.  The data we present are obtained with mass $am=0.0025$, and in some cases we consider smaller values to check for finite mass effects.  Since we would like to see if and how the scaling exponent changes with $\lambda$, we fit the mode number over a range $\Delta \lambda$, as
\begin{equation}
  \label{eq:nu_fitting}
  \nu(\lambda + \Delta\lambda/2) - \nu(\lambda-\Delta\lambda/2) = c\left[(\lambda+\Delta\lambda/2)^{\alpha+1} - (\lambda-\Delta\lambda/2)^{\alpha+1} \right].
\end{equation}
We carry out this two-parameter fit for fixed volume, mass and gauge coupling; it is fairly stable if $\Delta\lambda$ is not too small.
%%%%%%%%%%%%%%%%%%%%%%%%%%%%%%%%%%
\begin{figure}[tb]
  \centering
  \includegraphics[width=0.45\linewidth]{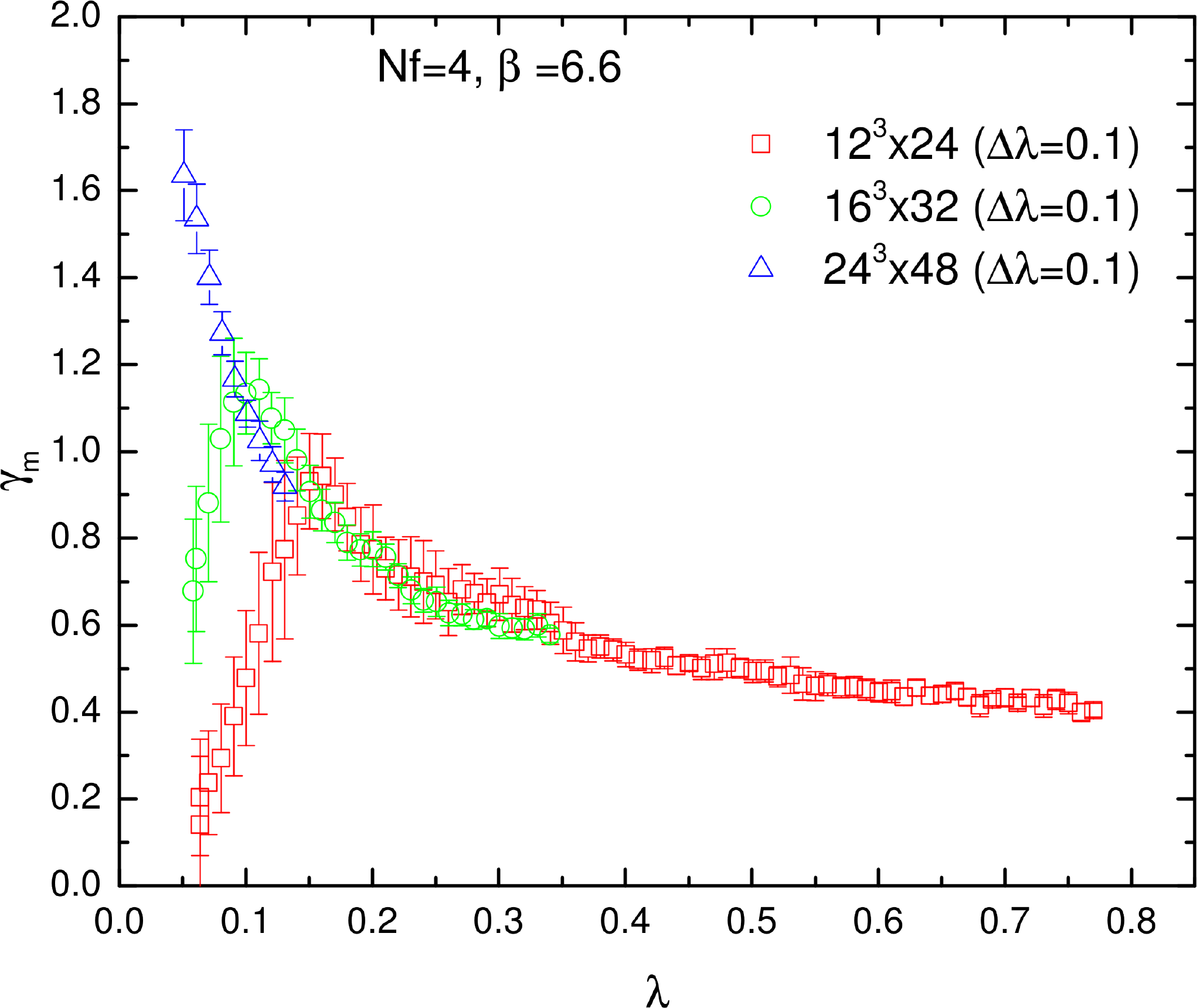}\hfill\includegraphics[width=0.45\linewidth]{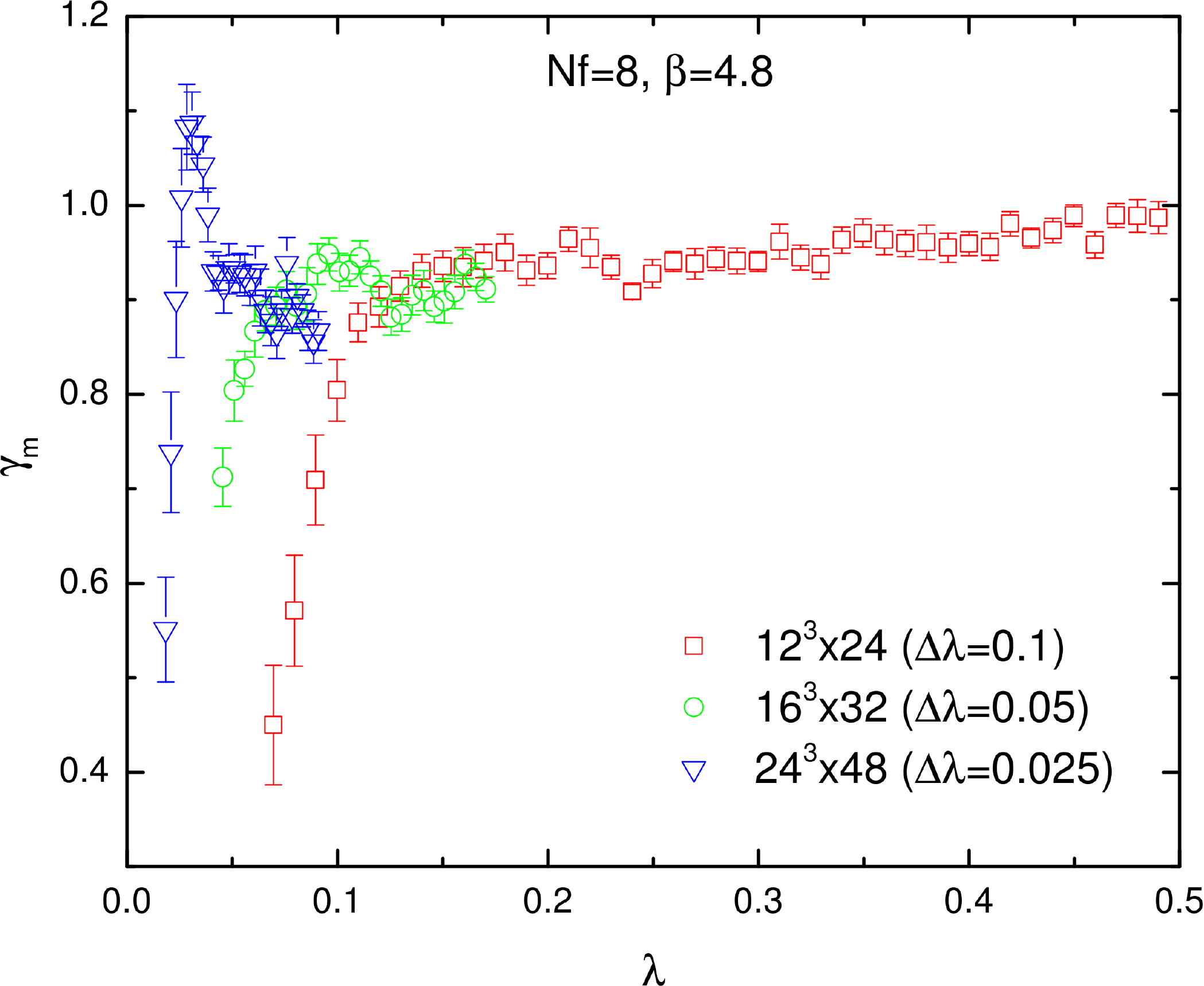}
  \caption{The anomalous dimension $\gamma_m$ based on the scaling relation of Eq.~\protect\ref{eq:nu_scaling} as a function of $\lambda$, on volumes $12^3 \X 24$ (red squares), $16^3 \X 32$ (green circles) and $24^3 \X48$ (blue triangles).  Left panel: $N_f=4$ flavor system at $\beta_F=6.6$, $am=0.0025$.  Right panel: $N_f=8$ flavor system at $\beta_F=4.8$, $am=0.0025$.
  \label{fig:rho_vs_lambda-1}}
\end{figure}
%%%%%%%%%%%%%%%%%%%%%%%%%%%%%%%%%

The left panel of \fig{fig:rho_vs_lambda-1} shows the anomalous dimension $\gamma_m=[4/(\alpha+1)-1]$ as a function of $\lambda$ in the 4 flavor system at $\beta_F=6.6$.  At this coupling and mass all three volumes are chirally symmetric, $\rho(0) = 0$.  The transient finite volume effects at small $\lambda$ are clear, but as $\lambda$ increases the predictions from all volumes fall onto the same curve.  The interesting feature of this plot is the steady decrease of $\gamma_m$ with increasing $\lambda$, i.e.\ towards the ultraviolet. This is the qualitative energy dependence expected from the perturbative relation $\gamma_m = 6 C_F g^2/(4\pi)^2$.  With $g^2(\lambda)$ determined from the one-loop beta function, this relation gives a good overall description of the data, though with coefficients that differ from their one-loop values by a factor of three.  At small $\lambda$, $\gamma_m \sim 1$, and indeed at this coupling the $24^3\X48$ system appears almost chirally broken.

The right panel of \fig{fig:rho_vs_lambda-1} shows the $\lambda$ dependence of the anomalous dimension for the 8 flavor system at $\beta_F=4.8$. At $am=0.0025$ all three investigated volumes are chirally symmetric, with $\rho(0) = 0$.  Unlike the $N_f=4$ case, the common $\gamma_m$ at which all three volumes eventually settle shows very little dependence on $\lambda$.  As expected, the 8-flavor anomalous dimension runs much more slowly than for $N_f = 4$.  We find similar behavior at other bare couplings, though the value of $\gamma_m$ decreases steadily as $\beta_F$ moves to weaker coupling.  We are currently calculating the running coupling in this system, which will allow us to combine results at several $\beta_F$ into a function $\gamma_m(g^2)$ that we can compare to perturbation theory. At this point we can only conclude that the 8 flavor system exhibits slow running compared to $N_f = 4$, with large $\gamma_m \sim 1$ at $\beta_F=4.8$; we cannot yet quantify the energy range where this behavior persists.

%%%%%%%%%%%%%%%%%%%%%%%%%%%%%%%%%%
\begin{figure}[htb]
  \centering
  \includegraphics[width=0.45\linewidth]{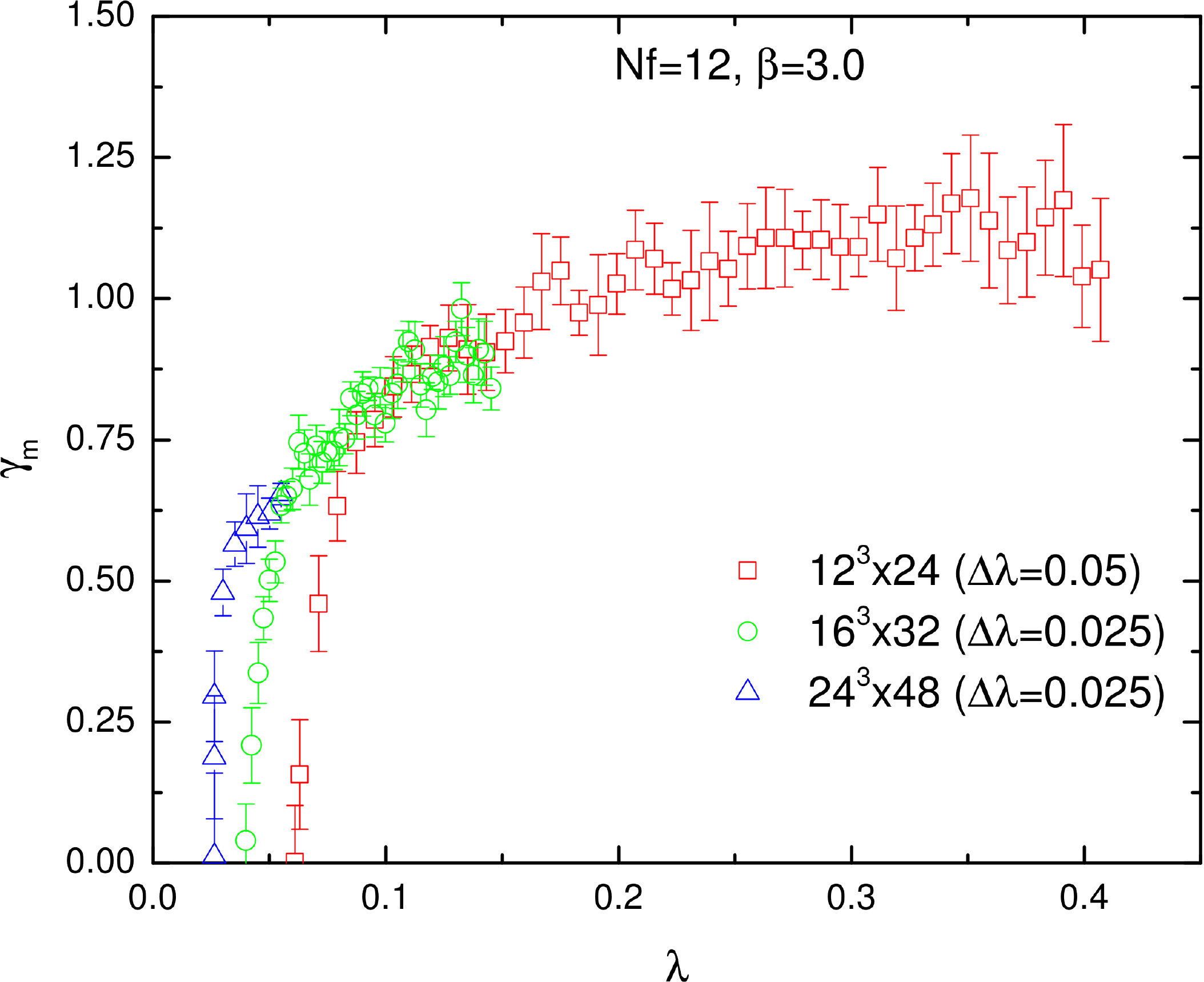}\hfill\includegraphics[width=0.45\linewidth]{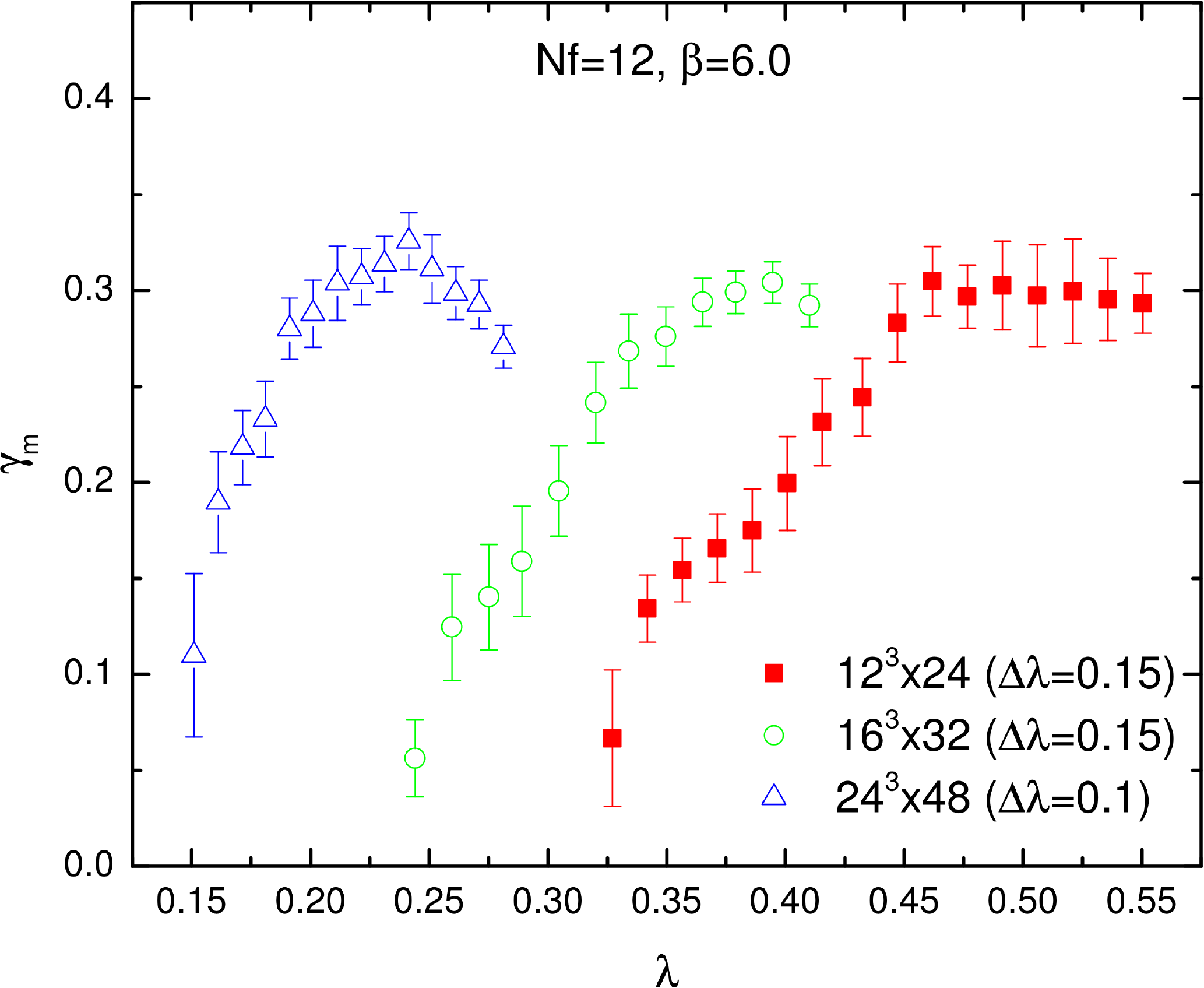}
  \caption{\label{fig:rho_vs_lambda-2} The anomalous dimension $\gamma_m$ based on the scaling relation of Eq.~\protect\ref{eq:nu_scaling} as a function of $\lambda$ for the $N_f=12$ flavor system on volumes $12^3 \X 24$ (red squares), $16^3 \X 32$ (green circles) and $24^3 \X48$ (blue triangles).  Left panel: $\beta_F=3.0$, $am=0.0025$.  Right panel: $\beta_F=6.0$, $am=0.0025$.}
\end{figure}
%%%%%%%%%%%%%%%%%%%%%%%%%%%%%%%%%

\fig{fig:rho_vs_lambda-2} shows two representative samples from our $N_f=12$ investigations. The left panel is the anomalous dimension at $\beta_F=3.0$, a fairly strong coupling but still on the weak coupling side of the \Sb phase.
Even at this fairly strong coupling, all three volumes are chirally symmetric, and they predict a consistent $\gamma_m$ at $\lambda$ large enough for finite volume effects to die out. However, the energy dependence is the opposite of what we observed with $N_f=4$: $\gamma_m$ increases with increasing $\lambda$. This is not consistent with the perturbative behavior at the $g^2=0$ Gaussian fixed point, and we interpret this result as indicating that the coupling $\beta_F=3.0$ is in the strong coupling side of the basin of attraction of the conformal infrared fixed point.

At weaker bare coupling (larger $\beta_F$) we observe less energy dependence, as the right panel of \fig{fig:rho_vs_lambda-2} illustrates at $\beta_F=6.0$.  The finite volume effects are more significant at this coupling, but the three volumes settle at a consistent value $\gamma_m \approx 0.3$ that is largely $\lambda$-independent. We find similar behavior at other $\beta_F \ge 6$.  Considering $6\le\beta_F\le8$, we determine the preliminary result $\gamma_m^* = 0.27(3)$ for the mass anomalous dimension at the IR fixed point.  We hope to obtain a more precise result for $\gamma_m^*$ in the future by performing a global fit over all the volumes, $\beta_F$ and ranges of $\lambda$ that produce consistent predictions.

While our result $\gamma_m^* = 0.27(3)$ is significantly smaller than that obtained from finite-size scaling by \refcite{Aoki:2012eq}, it is comparable to some of the results reported by \refcite{Fodor:2012uu}, which found different $0.2 \lesssim \gamma_m^* \lesssim 0.4$ depending on the observable used in the finite-size scaling analysis.  We also reported a larger value from investigations at stronger couplings~\cite{Cheng:2011ic}.  The fact that we need $\beta_F \geq 6$ to obtain scale-independent results could explain some of the inconsistencies at stronger couplings.

\section{Conclusion}

We presented preliminary results from our analysis of the mode number of the $N_f=4$, 8 and 12 flavor SU(3) models. We argued that a simple scaling form that depends on the mass anomalous dimension describes the mode number in an intermediate spectral range even for chirally broken systems. Our data show the running of $\gamma_m$ with the energy in the 4 flavor system, while with 8 flavors $\gamma_m$ does not change significantly over the accessible range of scales.  Our 12-flavor data are consistent with the existence of an infrared fixed point and show backward running at strong bare coupling.  At weaker couplings we obtain a preliminary prediction $\gamma_m^*=0.27(3)$ for the $N_f = 12$ mass anomalous dimension at the IR fixed point.

The Dirac eigenmodes offer an alternative method to study the infrared dynamics of strongly coupled systems. Our analysis at present is limited by the number of eigenvalues we can calculate on the larger volumes. The stochastic estimator proposed in \refcite{Giusti:2008vb} and used by \refcite{Patella:2012da} may greatly increase the reach of this approach.

% ------------------------------------------------------------------
\section*{Acknowledgments} % Draft complete
We thank Tamas Kovacs, Julius Kuti and Agostino Patella for helpful comments and discussions. There are many publications on the infrared dynamics of gauge-fermion systems with many flavors or higher representation fermions both from lattice and continuum studies that we could not refer to.  We apologize for the omissions.
This research was partially supported by the U.S.~Department of Energy (DOE) through Grant No.~DE-FG02-04ER41290 (A.~C., A.~H.\ and D.~S.) and by the DOE Office of Science Graduate Fellowship Program
%made possible in part by the American Recovery and Reinvestment Act of 2009, administered by ORISE-ORAU
under Contract No.~DE-AC05-06OR23100 (G.~P.).
Our code is based in part on the MILC Collaboration's public lattice gauge theory software,\footnote{\texttt{http://www.physics.utah.edu/$\sim$detar/milc/}} and on the PReconditioned Iterative MultiMethod Eigensolver (PRIMME) package~\cite{Stathopoulos:2010}.
Numerical calculations were carried out on the HEP-TH and Janus clusters at the University of Colorado; at Fermilab under the auspices of USQCD supported by the DOE SciDAC program; and at the San Diego Computing Center through the Extreme Science and Engineering Discovery Environment supported by National Science Foundation Grant No.~OCI-1053575.
% ------------------------------------------------------------------

{\renewcommand{\baselinestretch}{0.86}
 \bibliography{eigenmodes}

\providecommand{\href}[2]{#2}\begingroup\raggedright\begin{thebibliography}{10}

\bibitem{Cheng:2011ic}
A.~Cheng, A.~Hasenfratz and D.~Schaich, {\em Phys. Rev.} {\bf D85} (2012) 094509
  [\href{http://arXiv.org/abs/1111.2317}{{\tt 1111.2317}}].

\bibitem{Schaich:2012lat12}
D.~Schaich, A.~Cheng, A.~Hasenfratz and G.~Petropoulos, {\em PoS} {\bf Lattice
  2012} (2012, to appear).

\bibitem{Petropoulos:2012Lat12}
G.~Petropoulos, A.~Cheng, A.~Hasenfratz and D.~Schaich, {\em PoS} {\bf Lattice
  2012} (2012, in preparation).

\bibitem{Hasenfratz:2011xn}
A.~Hasenfratz, {\em Phys. Rev. Lett.} {\bf 108} (2012) 061601
  [\href{http://arXiv.org/abs/1106.5293}{{\tt 1106.5293}}].

\bibitem{Fodor:2012uu}
Z.~Fodor, K.~Holland, J.~Kuti, D.~Nogradi, C.~Schroeder and C.~H. Wong, {\em
  PoS} {\bf Lattice 2011} (2012) 073
  [\href{http://arXiv.org/abs/1205.1878}{{\tt 1205.1878}}].

\bibitem{Aoki:2012eq}
Y.~Aoki, T.~Aoyama, M.~Kurachi, T.~Maskawa, K.-i. Nagai, H.~Ohki, A.~Shibata,
  K.~Yamawaki and T.~Yamazaki, \href{http://arXiv.org/abs/1207.3060}{{\tt
  1207.3060}}.

\bibitem{Deuzeman:2012pv}
A.~Deuzeman, M.~P. Lombardo and E.~Pallante, {\em PoS} {\bf Lattice 2011} (2012)
  083 [\href{http://arXiv.org/abs/1201.1863}{{\tt 1201.1863}}].

\bibitem{Lin:2012iw}
C.-J.~D. Lin, K.~Ogawa, H.~Ohki and E.~Shintani,
  \href{http://arXiv.org/abs/1205.6076}{{\tt 1205.6076}}.

\bibitem{Neil:2012cb}
E.~T. Neil, {\em PoS} {\bf Lattice 2011} (2011) 009
  [\href{http://arXiv.org/abs/1205.4706}{{\tt 1205.4706}}].

\bibitem{DeGrand:2009hu}
T.~DeGrand, {\em Phys. Rev.} {\bf D80} (2009) 114507
  [\href{http://arXiv.org/abs/0910.3072}{{\tt 0910.3072}}].

\bibitem{Giusti:2008vb}
L.~Giusti and M.~Luscher, {\em JHEP} {\bf 0903} (2009) 013
  [\href{http://arXiv.org/abs/0812.3638}{{\tt 0812.3638}}].

\bibitem{Patella:2012da}
A.~Patella, \href{http://arXiv.org/abs/1204.4432}{{\tt 1204.4432}}.

\bibitem{Banks:1979yr}
T.~Banks and A.~Casher, {\em Nucl. Phys.} {\bf B169} (1980) 103.

\bibitem{Stathopoulos:2010}
A.~Stathopoulos and J.~R. McCombs, {\em ACM Trans. Math. Softw.} {\bf 37} (2010)
  21.

\end{thebibliography}\endgroup
 \bibliographystyle{JHEP-2}}
\end{document}